\documentclass[
    ,final            
  ]
  {aipproc}

\layoutstyle{8x11double}

\usepackage{amssymb,amsmath,amstext,amsthm,amsfonts}

\def\be{\begin{equation}}
\def\ee{\end{equation}}
\def\bea{\begin{eqnarray}}
\def\eea{\end{eqnarray}}
\def\bear{\begin{array}}
\def\ear{\end{array}}
\def\bfig{\begin{figure}}
\def\efig{\end{figure}}
\def\bcen{\begin{center}}
\def\ecen{\end{center}}
\def\raw{\rightarrow}

\def\bi{\begin{itemize}}
\def\ei{\end{itemize}}

\begin{document}

\title{Theoretical highlights of neutrino-nucleus interactions}

\classification{25.30.-c, 24.10.-i, 13.15.+g, 14.20.Gk, 13.40.Gp}
\keywords      {electron- and neutrino-nucleus reactions, form factors, baryon resonances, quasielastic scattering, pion production}

\author{Luis Alvarez-Ruso}{
  address={Centro de F\'isica Computacional, Departamento de F\'isica, Universidade de Coimbra, Portugal}
}

\begin{abstract}
The recent theoretical developments in the field of neutrino-nucleus interactions in the few-GeV region are reviewed based on the presentations made at the NuInt09 Workshop. The topics of electron scattering and its connections with neutrino interactions, neutrino induced quasielastic scattering and pion production (coherent and incoherent) are covered, with special emphasis on the challenges that arise in the comparison with new experimental data.  
\end{abstract}

\maketitle


\section{Introduction}

Neutrino interactions with nuclei have received a considerable attention in recent years stimulated by the needs of neutrino oscillation experiments. A variety of theoretical calculations have been performed for the different reaction channels. At the same time, new high quality data are becoming available from MiniBooNE, MINOS, NOMAD and SciBooNE, and more is expected from MINER$\nu$A, an experiment fully dedicated to cross section (CS) measurements. These problems have been addressed in the Sixth International Workshop on Neutrino-Nucleus Interactions in the Few-GeV Region (NuInt09)~\cite{web}. In this article, I present some of the theoretical highlights of this meeting, focusing on the most relevant reaction channels for accelerator-based experiments: quasielastic (QE) scattering and pion production ($\pi$P); electron-nucleus 
($e A$) interactions and their relevance for neutrino-nucleus ($\nu A$) scattering are also discussed. To begin, the motivations for doing theoretical and experimental research on $\nu A$ CS are reviewed. 

\section{Motivation}

{\bf Oscillation experiments}~\cite{Tanaka@NuInt09}.
Nowadays, the main reason for CS studies is the demand from oscillation experiments. Next generation ones aim at a precise determination of mass-squared difference $\Delta m_{32}$ and mixing angle $\theta_{23}$ in $\nu_\mu$ disappearance measurements. The ability to reconstruct the neutrino energy is crucial for this program. Indeed, the oscillation probability
$
P(\nu_\mu \raw \nu_\tau) = \sin^2{2 \theta_{23}} \, \sin^2{[\Delta m^2_{32} L/( 2 E_\nu)] }
$ 
depends on the neutrino energy $E_\nu$, which is not known for broad fluxes. The neutrino energy is usually reconstructed from charged-current quasielastic scattering (CCQE) events  $\nu_\mu \, n \raw \mu^- \, p$, dominant at $E_\nu \sim 1$~GeV. If the energy and angle of the final lepton are measured, $E_\nu$ can be determined with two-body kinematics, but this procedure
is exact only for free neutrons. As detectors are composed of nuclei, the reconstructed energy is smeared due to the momentum distribution of the bound nucleons. Moreover, the $E_\nu$ determination could be wrong for a fraction of events that are not CCQE ones but look identical to them in the detector. These are mainly $\Delta (1232)$ excitation events where the $\Delta$ is absorbed before decaying. Rejecting them requires a good command of final state interactions (FSI).   

High sensitivity searches for $\nu_\mu \raw \nu_e$ appearance, associated with $\theta_{13}$ and CP violation,  also rely on good CS knowledge. The main background is neutral current (NC) $\pi^0$ production because the electromagnetic showers from pion decay photons can be misidentified as electrons from $\nu_e$ induced CCQE $\nu_e \, n \raw e^- \, p$ if one of the photons is not identified. 

{\bf Astrophysics}~\cite{Balantekin:2009qq}. 
Neutrinos play an important role in astrophysical phenomena and carry information about the emitting sources. More precise measurements of the CNO neutrinos will provide a test for energy generation in stars and a better understanding of solar metalicity. The dynamics of core-collapse supernovae is controlled by neutrino interactions. The neutron rich environment of supernovae is a candidate site for r-process nucleosynthesis because radiated neutrinos convert neutrons into protons. These questions require a good knowledge of low energy neutrino production and detection CS. Although some neutrino CS of astrophysical interest could be investigated with beta beams, most of them cannot be measured and one relies on shell model or random phase approximation (RPA) calculations.   

{\bf Physics beyond standard model}~\cite{Balantekin:2009qq, Ohlsson@NuInt09}. 
Neutrino CS might be used to set bounds on nonstandard neutrino interactions. For example, deviations from universality in the $Z\nu\nu$ vertex could be accessed in deep inelastic scattering experiments at TeV energies. At low energies, a measurement of weak magnetism in  $\bar{\nu}_e \, p \raw e^+ \, n$ may provide another test of vector current conservation. 

{\bf Hadronic physics}.  
With high intensity neutrino beams it is possible to investigate the axial structure of the nucleon and baryon resonances. This information shall enlarge our view of hadron structure beyond what is presently known about electromagnetic form factors (FF) from JLab. MINER$\nu$A will probe the four-momentum transfer squared ($Q^2$) dependence of the nucleon axial FF with unprecedented accuracy and also study neutrino induced $\pi$P, which is dominated by resonance excitation. Another fundamental question is the strangeness content of the nucleon spin which can be best unraveled with NCQE $\nu \, p(n) \raw \nu \, p(n)$ reactions. 

{\bf Nuclear physics}. 
Modern neutrino experiments are performed with nuclear targets.  For nuclear physics this represents a challenge and an opportunity. A challenge because the above-mentioned prospects for more precise knowledge of neutrino and baryon properties require that nuclear effects are under control. An opportunity because $\nu A$ CS incorporate a richer information than $e A$ ones providing an excellent testing ground for nuclear structure, many-body mechanisms and reaction models.  

\section{Electron scattering}

There are many theoretical methods that can be applied to $\nu A$ interactions 
depending on the kinematic region: from shell model at low energies, through Fermi gas with hadronic degrees of freedom at intermediate energies, to perturbative QCD at the highest ones. All of them have been applied to $e A$ scattering. Moreover, some have been developed with the aim of understanding the large amount of 
good quality data acquired in decades of experimental studies. A good description of electron scattering data is mandatory for neutrino interaction models. 

In addition, electron scattering data on proton and deuteron are used to extract electromagnetic nucleon elastic and transition ($N-\Delta$, $N-N^*$) FF, an input for the weak ($V-A$) hadronic currents since the $V$ FF can be expressed in terms of the electromagnetic ones assuming isospin symmetry. Thanks to JLab data it has been established that proton electric and magnetic FF do not have the same behavior at large $Q^2$~\cite{Jones:1999rz}. These developments are incorporated in the FF parametrizations provided by Bodek et al.~\cite{Bodek:2007vi}. For resonance excitation, a unitary isobar model (MAID) has been used to extract the transition helicity amplitudes from the world data on pion photo- and electro-production for all four star resonances with masses below 2~GeV~\cite{Drechsel:2007if}. The helicity amplitudes can be mapped into the electromagnetic FF. The analysis reveals that the $N-\Delta(1232)$ transition is not purely magnetic. This has some impact on neutrino induced $\pi$P CS~\cite{Leitner:2008ue}. 

Let us consider now inclusive $e A$ scattering, where only the final electron is detected. The simplest model for this reaction in the QE region, and the one implemented by default in all event generators used in the analysis of neutrino data, is the relativistic global Fermi gas (FG) model. It assumes that the interaction takes place on single nucleons whose contributions are summed incoherently [impulse approximation (IA)]. The struck nucleons have momentum distributions characterized by a Fermi momentum $p_F$, and a constant binding energy $\epsilon_B$. Outgoing nucleons cannot go into occupied states (Pauli blocking)~\cite{Smith:1972xh}. With two parameters $(p_F,\epsilon_B)$, the main features of the inclusive QE CS  can be explained. However, a more careful view reveals the shortcomes of this simple picture. In particular, the global FG model overestimates the longitudinal response. Nuclear dynamics must be taken into account~\cite{Benhar@NuInt09}.    

The presence of nucleon-nucleon (NN) interactions implies that nucleon propagators are dressed with complex selfenergies $\Sigma$. In other words, the nucleons do not have a well defined dispersion relation but become broad states characterized by spectral functions 
\be
S_{h,p}(p) = - \frac{1}{\pi} \frac{\mathrm{Im}  \Sigma(p)}{[p^2-m_N^2-\mathrm{Re} \Sigma(p)]^2 + [\mathrm{Im}  \Sigma(p)]^2} \,,
\ee  
for both struck (holes) and outgoing (particles) nucleons. $S_h$ includes an 80-90\% contibution from single-particle states while the rest of the nucleons participate in NN interactions (correlations) and are located at high momentum (see Fig.~4 of Ref.~\cite{Ankowski:2007uy}). $S_p$ includes the effect of the interaction of the outgoing nucleon with the medium, often accounted for with an optical potential (OP) within the Glauber approximation~\cite{Ankowski:2007uy}. An alternative approach~\cite{Leitner:2008ue} adopts a local FG model ($p_F (r) = [(3/2) \pi^2 \rho(r)]^{1/3}$ with $\rho(r)$ the nuclear density), which introduces  space-momentum correlations absent in the global FG (see Fig.~6 of Ref.~\cite{Leitner:2008ue}). Instead of the constant binding, all  nucleons are exposed to a density and momentum dependent mean field potential. $S_p$ is considered in full glory, employing the low density approximation to calculate $\mathrm{Im} \Sigma$, but it is argued that the correlated part of $S_h$ plays a minor role in the description of inclusive CS and is neglected. With both frameworks a good description of inclusive data in the QE region is achieved (see Figs.~2,3 of Ref.~\cite{Nakamura:2007pj} and Figs.~9,10 of Ref.~\cite{Leitner:2008ue}) improving the result of the global FG. Furthermore, they can be extended to include resonances and nonresonant $\pi$P. While the agreement with data at the second ($\Delta$) peak is also good, the description of the dip region between the QE and $\Delta$ peaks requires 2-particle-2-hole ($2p-2h$) excitations from meson exchange currents (MEC)~\cite{Gil:1997bm}. The dip region is important for $\nu A$ experiments because many CCQE-like events originate there.

There is another class of IA relativistic models originally developed for QE electron scattering and later extended to neutrino scattering~\cite{Giusti:2009im,Udias@NuInt09}.  The initial nucleons are treated as single-particle bound states whose wave functions are solutions of the Dirac equation with a $\sigma$-$\omega$ mean field potential. The various treatments of the final state include plane-wave IA, where the interaction of the outgoing nucleon with the medium is neglected, the so called relativistic mean field model (RMF), where the scattering wave functions are calculated with the same energy independent real potential used for the bound nucleons, and distorted-wave IA with complex OP (DWIA); in this case the scattering states are obtained by solving the Dirac equation or using the Glauber model. DWIA models are successful in describing a large amount of exclusive proton knockout $(e,e' p)$ data but are not appropriate for inclusive scattering. The imaginary part of the OP produces an absorption and a reduction of the CS (see Fig.~1 of Ref.~\cite{Giusti:2009im}), which accounts for the flux lost towards other channels. This is correct for an exclusive reaction but not for an inclusive one where all channels contribute and the total flux must be conserved. This approximation that retains only the real part for inclusive processes conserves the flux but is conceptually wrong because the OP has to be complex owing to the presence of inelastic channels~\cite{Giusti:2009im}. An alternative is the Green function approach~\cite{Meucci:2003uy} where the imaginary part of the OP is responsible for the flux redistribution among different channels. This method guarantees a consistent treatment of both exclusive and inclusive reactions. An excellent description of the QE peak at different energies is achieved (Fig.~8 of Ref.~\cite{Meucci:2003uy}) although the transverse response is overestimated due of the lack of more complicated mechanisms such as MEC.

Inclusive electron scattering data exhibit interesting systematics that can be used to predict $\nu A$ CS. When the  experimental $(e,e')$ differential CS are divided by the corresponding single nucleon CS and multiplied by the global Fermi momentum, the resulting function 
\be
f= p_F \, \frac{\frac{d\sigma}{d\Omega d\omega}}{Z \sigma_{ep} + N \sigma_{en}}
\ee
is found to depend on energy and 3-momentum transfers $(\omega, |\vec{q}|)$ through a particular combination, the scaling variable $\psi'$, and to be largely independent of the specific nucleus  (superscaling)~\cite{Barbaro:2009rx}. Scaling violations reside mainly in the transverse channel. Therefore, an experimental scaling function $f(\psi')$ could be extracted by fitting the data for the longitudinal response. The experimental $f(\psi')$ has an asymmetric shape with a tail at positive $\psi'$ (large $\omega$). The requirement of a realistic description of the scaling function is a constraint for nuclear models. The relativistic FG model with exact superscaling gives a wrong symmetric shape for $f(\psi')$ while the RMF model reproduces it well. With the superscaling approximation (SuSA) a good representation of the nuclear response can be obtained by embedding nuclear effects in the scaling function: the observables can be calculated with the simple relativistic FG model followed by the replacement $f_{FG} \raw f_{exp}$. The same strategy can be used to predict $\nu A$ CS, minimizing the model dependence of the results. SuSA predicts 15~\% smaller total CCQE CS compared to the relativistic FG. (Fig.~3 of Ref.~\cite{Amaro:2006tf}). It should be remembered that scaling fails at $\omega < 40$~MeV and $|\vec{q}| < 400$~MeV due to collective effects.  

\section{Neutrino induced QE scattering}

The CCQE scattering amplitude on a single nucleon is proportional to the product of the leptonic and hadronic currents. The hadronic one is given in terms of vector and axial FF $F_{A,P}$. $F_P$ can be related to $F_A$ using PCAC. $F_A$ is usually parametrized as  $F_A (Q^2) = g_A \left( 1 + Q^2/M_A^2 \right)^{-2}$, with $g_A=1.27$ from $\beta$ decay. For $M_A$, the world average value from early experiments is $M_A = 1.026 \pm 0.021$~GeV~\cite{Liesenfeld:1999mv} and the one extracted from threshold $\pi$ electroproduction data $\tilde{M}_A = 1.069 \pm 0.016$~GeV~\cite{Liesenfeld:1999mv}. While the result of early neutrino experiments might be questionable due to the low statistics and poor knowledge of the neutrino flux, $\pi$ electroproduction offers a solid indication that, at least at low $Q^2$, $M_A \sim 1$~GeV. 

Modern experiments  have started to provide a wealth of data on neutrino induced QE scattering for different energies and nuclear targets.  MiniBooNE, running at $\langle E_\nu \rangle \sim 750$~MeV on a CH$_2$ target, has collected the largest sample available so far for low energy $\nu_\mu$ CCQE~\cite{:2007ru}. After subtracting the non CCQE background using NUANCE~\cite{Casper:2002sd}, the CCQE data set was analyzed with the relativistic global FG model. The shape of the muon angular and energy distributions averaged over the $\nu_\mu$ flux could be described with standard values of $p_F$ and $\epsilon_B$, but restricting the phase space for the final proton by means of an ad hoc parameter $\kappa = 1.019 \pm 0.011$,  and taking $M_A = 1.23 \pm 0.20$~GeV~\cite{:2007ru}. This value of $M_A$ is considerably higher than the world average and the NOMAD result at high energies (3-100~GeV), also on $^{12}$C [$M_A = 1.05 \pm 0.02(stat) \pm 0.06(syst)$~GeV]~\cite{Lyubushkin:2008pe}. A recent MiniBooNE reanalysis, using CC single $\pi$P data to adjust the simulation employed to subtract the background, obtains $\kappa = 1.007 \pm 0.007$ and an even higher $M_A = 1.35 \pm 0.17$~GeV~\cite{Katori:2009du}. 

The introduction of $\kappa$ offers a convenient way of parametrizing the low $Q^2$ reduction shown by the data within a simple FG model but its physical meaning is obscure. As explained above, the high $M_A$ is also hard to understand. An alternative is that the observed increase of the CS at higher $Q^2$ might reflect the underlying nuclear physics rather than the nucleon's. A more realistic description using spectral functions  explains neither the low $Q^2$ reduction nor the high $Q^2$ increase~\cite{Benhar:2009wi,AlvarezRuso:2009ad}. In Ref.~\cite{AlvarezRuso:2009ad}, it has been shown that the shape of the $Q^2$ distribution extracted from data by MiniBooNE could be reproduced fairly well with $M_A =1$~GeV by taking into account the renormalization of the electroweak couplings caused by the presence of strongly interacting nucleons (long range RPA correlations). This phenomenon, known as quenching, is well established in nuclear $\beta$ decay and has proved crucial for a simultaneous description of muon capture on $^{12}$C and the low energy LSND CCQE measurement~\cite{Nieves:2004wx}.  The problem is that quenching causes a reduction of the integrated CS already with respect to the FG result with $M_A =1$~GeV. The model prediction $\langle \sigma \rangle = 3.2 \times 10^{-38}$~cm$^2$ is much smaller than MiniBooNE's result of $5.65 \times 10^{-38}$~cm$^2$ with an error of 10.8~\%~\cite{Katori:2009du}. Other approaches like SUSA and RMF also find a reduction in the integrated CS. A promising solution to this puzzle has been proposed in a recent article~\cite{Martini:2009uj}, where the additional strength measured by MiniBooNE is explained by the contribution of $2p-2h$ states that are not experimentally distinguishable from the standard $1p-1h$ CCQE. More accurate comparisons to data (preferably inclusive) are still required: although the authors of Ref.~\cite{Casper:2002sd} stress that the dominant contributions to the $2p-2h$ CS do not reduce to a modification of the $\Delta$ width, the background subtracted from the MiniBooNE data includes a supression of $\pi$ production that, in principle, accounts for pionless resonance decay~\cite{Casper:2002sd}.  

\section{Pion production} 

Pion production in nuclei can be incoherent if the final nucleus is excited, or coherent if the nucleus remains in its ground state. The first step towards a good description of $\pi$P on nuclear targets is a realistic model of the elementary reaction (on nucleons). In the few-GeV region, the $\Delta(1232)$ excitation, followed by $\Delta \raw \pi N$ is the dominant mechanism. Little is known about the axial $N-\Delta$ FF, often denoted as $C^A_{3-6}$. At moderate $Q^2$, $C^A_5$ is the relevant one. The value of $C^A_5 (0)$ can be related to the $\Delta N \pi$ coupling: $C^A_5 (0) = g_{\Delta N \pi} f_\pi / (\sqrt{6} \, m_N) \approx 1.2$. The $Q^2$ behavior can be obtained by comparison to deuterium data from ANL and BNL bubble chamber experiments, with good statistics but large systematic errors due to the poor knowledge of the neutrino flux. In addition, there are nonresonant contributions that, close to threshold, are fully determined by chiral symmetry~\cite{Hernandez:2007qq}. The inclusion of these nonresonant terms led the authors of Ref.~\cite{Hernandez:2007qq} to reduce $C^A_5 (0)= 0.867 \pm 0.075$ to describe the data, but the fit was done to ANL data alone, which are systematically below BNL ones. A combined analysis of both ANL and BNL data taking into account normalization uncertainties and deuteron effects, but not the nonresonant background, obtains $C^A_5 (0) = 1.19 \pm 0.08$~\cite{Graczyk:2009qm}. Another theoretical description is based on a dynamical model of photo-, electro- and weak $\pi$P~\cite{Sato:2003rq}. Starting from an effective Hamiltonian with $N-\Delta$ couplings obtained with the constituent quark model ($\sim 30$~\% below the measured ones), the $T$ matrix is obtained by solving the Lippmann-Schwinger equation in coupled channels. In this way the bare couplings get renormalized by meson clouds. The predicted CS are in good agreement with data (Figs. 5-8 of Ref.~\cite{Sato:2003rq}).  

When $\pi$P takes place inside the nucleus, the elementary cross section is modified. The most important in-medium change in this case is the modification of the $\Delta$, whose mass gets shifted and its width increased due to absorption processes, mainly $\Delta \, N \raw N \, N$. The produced pion interacts strongly with the nuclear environment: it can be absorbed or scatter with the nucleons with and without charge exchange. At intermediate energies, a large number of states can be excited, so the description of the final system, in particular pion and nucleon propagation, requires a semiclassical treatment. The Giessen Boltzmann-Uehling-Uhlenberg (GiBUU) model allows to study these FSI in a realistic manner using transport theory in coupled channels.  The effect of FSI on pion spectra appears to be large, especially for heavy targets (see for instance Figs.~13 and 14 of Ref.~\cite{Leitner:2006ww}). The ratio  $\sigma(CC1\pi^+)/\sigma(CCQE)$ as a function of the neutrino energy has also been studied and compared to the observed (i.e. without FSI corrections) MiniBooNE experimental result~\cite{AguilarArevalo:2009eb}. The calculation should include all CCQE-like events in the denominator. It is found that the theory clearly underestimates the data at $E_\nu > 1$~GeV (see Fig.~3 of Ref.~\cite{Leitner:2009ec}). A similar result is obtained in Ref.~\cite{Athar:2009rc} with a model that accounts for in-medium modifications of the elementary CS and propagates the pions through the medium with a cascade. Reference~\cite{Martini:2009uj} finds a good agreement for the observed ratio (bottom panel of Fig. 19) but without including pion FSI. The latter will reduce the numerator and increase the denominator spoiling the agreement to some extent.     

Coherent $\pi$P occurs at very low $Q^2$ where the nucleus is less likely to break. It has a very small CS compared to the incoherent process, but relatively larger than coherent $\pi$P induced by photons or electrons due to the non vanishing contribution of the axial current at the relevant kinematics~\cite{Amaro:2008hd}. This reaction has attracted the attention of theoreticians because the low energy experiments find CS smaller than predicted by the pioneering model of Rein and Sehgal (RS)~\cite{Rein:1982pf}. The RS model uses PCAC in the $Q^2=0$ limit to relate neutrino induced coherent $\pi$P to pion-nucleus ($\pi A$) elastic scattering, which is modeled in terms of the pion-nucleon CS. By taking the $Q^2=0$ limit, the RS model neglects important angular dependence at low energies~\cite{Hernandez:2009vm}. This, together with the fact that the description of the $\pi A$ elastic CS is not realistic~\cite{Hernandez:2009vm}, results in CS well above the experimental data. As the energy increases, these deficiencies become less relevant and the RS model is revealed as a simple and elegant method. An alternative approach based on PCAC directly uses the experimental $\pi A$ elastic CS~\cite{Berger:2008xs,Paschos:2009ag}. In this way the treatment of the outgoing pion is improved, but a spurious initial pion distortion, present in $\pi A$ elastic scattering but not in coherent $\pi$P, is introduced. With this method smaller and more compatible with the experiments CS are obtained. Microscopic approaches meant to work in the $\Delta$ region (and only there) have also been developed. They combine the $\Delta$ excitation picture of weak $\pi$P on the nucleon or the more complete models of Refs.~\cite{Hernandez:2007qq,Sato:2003rq} with the $\Delta$-hole model. As the nucleus remains in its ground state, a quantum treatment of pion distortion is feasible by means of the eikonal approximation~\cite{Singh:2006bm}, the Klein-Gordon~\cite{AlvarezRuso:2007it,Amaro:2008hd} or the Lippmann-Schwinger~\cite{Nakamura:2009iq} equations with realistic OP. It is found that medium effects and pion distortion reduce considerably the CS and shift the peak to lower pion-momenta (see for example Fig.~2 of Ref.~\cite{Amaro:2008hd}). Nonlocalities in the $\Delta$ propagation were found relevant~\cite{Leitner:2009ph} and have been explicitly taken into account in Ref.~\cite{Nakamura:2009iq} but not in Refs.~\cite{Singh:2006bm,AlvarezRuso:2007it,Amaro:2008hd}, where they might have been partially included via the empirical $\Delta$ mass shift~\cite{Nakamura:2009iq}. The coherent $\pi$P CS is very sensitive to the value of $C^A_5 (0)$~\cite{AlvarezRuso:2007it}. A preliminary comparison to MiniBooNE NC$\pi^0$ data~\cite{Anderson@NuInt09} indicates that a  $C^A_5 (0) \sim 1.2$ is preferred.

\section{Conclusions}
 
This is an excellent time for theoretical studies on $\nu A$ interactions as new high quality data have started to appear. A good understanding of (semi)inclusive $\nu A$ (together with $e A$) CS is required for the (model dependent) separation of different mechanisms: only then more precise determinations of $E_\nu$ and of the NC$\pi^0$ background will be possible. The theoretical progress should also find its way to the simulations employed by the experiments.


\begin{theacknowledgments}
I thank all the NuInt09 participants, whose presentations defined the current status of this field and the NuFact09 organizers for the invitation to present this summary. 

\end{theacknowledgments}



\begin{thebibliography}{19}

\bibitem{web} NuInt09 website, \url{http://nuint09.ifae.es}.

\bibitem{Tanaka@NuInt09} H. A. Tanaka, NuInt09 Proceedings (AIP Conf. Proc.).

\bibitem{Balantekin:2009qq}
  A.~B.~Balantekin, 
  NuInt09 Proceedings (AIP Conf. Proc.), arXiv:0909.0226 [hep-ph].

\bibitem{Ohlsson@NuInt09} T. Ohlsson, NuInt09 Proceedings (AIP Conf. Proc.).

\bibitem{Jones:1999rz}
  M.~K.~Jones {\it et al.},
  Phys.\ Rev.\ Lett.\  {\bf 84}, 1398 (2000).


\bibitem{Bodek:2007vi}
  A.~Bodek {\it et al.},
  J.\ Phys.\ Conf.\ Ser.\  {\bf 110}, 082004 (2008).  

\bibitem{Drechsel:2007if}
  D.~Drechsel, S.~S.~Kamalov and L.~Tiator,
  Eur.\ Phys.\ J.\  A {\bf 34}, 69 (2007).  

\bibitem{Leitner:2008ue}
  T.~Leitner {\it et al.},
  Phys.\ Rev.\  C {\bf 79}, 034601 (2009).  

\bibitem{Smith:1972xh}
  R.~A.~Smith and E.~J.~Moniz,
  Nucl.\ Phys.\  B {\bf 43}, 605 (1972)
  [Erratum-ibid.\  B {\bf 101}, 547 (1975)].

\bibitem{Benhar@NuInt09} O. Benhar, NuInt09 Proceedings (AIP Conf. Proc.).

\bibitem{Ankowski:2007uy}
  A.~M.~Ankowski and J.~T.~Sobczyk,
  Phys.\ Rev.\  C {\bf 77}, 044311 (2008).

\bibitem{Nakamura:2007pj}
  H.~Nakamura {\it et al.},
  Phys.\ Rev.\  C {\bf 76}, 065208 (2007).  

\bibitem{Gil:1997bm}
  A.~Gil, J.~Nieves and E.~Oset,
  Nucl.\ Phys.\  A {\bf 627}, 543 (1997).  

\bibitem{Giusti:2009im}
  C.~Giusti {\it et al.},
  NuInt09 Proceedings (AIP Conf. Proc.), arXiv:0910.1045 [nucl-th].

\bibitem{Udias@NuInt09} J. L. Herraiz {\it et al.}, NuInt09 Proceedings (AIP Conf. Proc.).

\bibitem{Meucci:2003uy}
  A.~Meucci {\it et al.},
  Phys.\ Rev.\  C {\bf 67}, 054601 (2003).  

\bibitem{Barbaro:2009rx}
  M.~B.~Barbaro {\it et al.},
  NuInt09 Proceedings (AIP Conf. Proc.), arXiv:0909.2602 [nucl-th].

\bibitem{Amaro:2006tf}
  J.~E.~Amaro {\it et al.},
  Phys.\ Rev.\ Lett.\  {\bf 98}, 242501 (2007).

\bibitem{Liesenfeld:1999mv}
  A.~Liesenfeld {\it et al.},
  Phys.\ Lett.\  B {\bf 468}, 20 (1999).  


\bibitem{:2007ru}
  A.~A.~Aguilar-Arevalo {\it et al.},
  Phys.\ Rev.\ Lett.\  {\bf 100}, 032301 (2008).

\bibitem{Casper:2002sd}
  D.~Casper,
  Nucl.\ Phys.\ Proc.\ Suppl.\  {\bf 112}, 161 (2002).
  
\bibitem{Lyubushkin:2008pe}
  V.~Lyubushkin {\it et al.},
  Eur.\ Phys.\ J.\  C {\bf 63}, 355 (2009)
.  

\bibitem{Katori:2009du}
  T.~Katori,
  NuInt09 Proceedings (AIP Conf. Proc.), arXiv:0909.1996 [hep-ex].

\bibitem{Benhar:2009wi}
  O.~Benhar and D.~Meloni,
  arXiv:0903.2329 [hep-ph].

\bibitem{AlvarezRuso:2009ad}
  L.~Alvarez-Ruso {\it et al.},
  NuInt09 Proceedings (AIP Conf. Proc.), arXiv:0909.5123 [nucl-th].
  

\bibitem{Nieves:2004wx}
  J.~Nieves, J.~E.~Amaro and M.~Valverde,
  Phys.\ Rev.\  C {\bf 70}, 055503 (2004)
  [Erratum-ibid.\  C {\bf 72}, 019902 (2005)].


\bibitem{Martini:2009uj}
  M.~Martini {\it et al.},
  arXiv:0910.2622 [nucl-th].

\bibitem{Hernandez:2007qq}
  E.~Hernandez, J.~Nieves and M.~Valverde,
  Phys.\ Rev.\  D {\bf 76}, 033005 (2007).  

\bibitem{Graczyk:2009qm}
  K.~M.~Graczyk {\it et al.},
  arXiv:0908.2175 [hep-ph].

\bibitem{Sato:2003rq}
  T.~Sato, D.~Uno and T.~S.~H.~Lee,
  Phys.\ Rev.\  C {\bf 67}, 065201 (2003).  

\bibitem{Leitner:2006ww}
  T.~Leitner, L.~Alvarez-Ruso and U.~Mosel,
  Phys.\ Rev.\  C {\bf 73}, 065502 (2006).

\bibitem{AguilarArevalo:2009eb}
  A.~A.~Aguilar-Arevalo {\it et al.},
  Phys.\ Rev.\ Lett.\  {\bf 103}, 081801 (2009).

\bibitem{Leitner:2009ec}
  T.~Leitner {\it et al.}, NuInt09 Proceedings (AIP Conf. Proc.),
    arXiv:0909.0838 [nucl-th].

\bibitem{Athar:2009rc}
  M.~S.~Athar, S.~Chauhan and S.~K.~Singh,
    arXiv:0908.1442 [nucl-th].

\bibitem{Amaro:2008hd}
  J.~E.~Amaro {\it et al.},
  Phys.\ Rev.\  D {\bf 79}, 013002 (2009).  


\bibitem{Rein:1982pf}
  D.~Rein and L.~M.~Sehgal,
  Nucl.\ Phys.\  B {\bf 223}, 29 (1983).

\bibitem{Hernandez:2009vm}
  E.~Hernandez, J.~Nieves and M.~J.~Vicente-Vacas,
  Phys.\ Rev.\  D {\bf 80}, 013003 (2009).  

\bibitem{Berger:2008xs}
  C.~Berger and L.~M.~Sehgal,
  Phys.\ Rev.\  D {\bf 79}, 053003 (2009).  

\bibitem{Paschos:2009ag}
  E.~A.~Paschos and D.~Schalla,
  Phys.\ Rev.\  D {\bf 80}, 033005 (2009).  

\bibitem{Singh:2006bm}
  S.~K.~Singh, M.~S.~Athar and S.~Ahmad,
  Phys.\ Rev.\ Lett.\  {\bf 96}, 241801 (2006).

\bibitem{AlvarezRuso:2007it}
  L.~Alvarez-Ruso, L.~S.~Geng and M.~J.~Vicente Vacas,
  Phys.\ Rev.\  C {\bf 76}, 068501 (2007)
  [Erratum-ibid.\  C {\bf 80}, 029904 (2009)].  

\bibitem{Nakamura:2009iq}
  S.~X.~Nakamura {\it et al.}, 
  arXiv:0910.1057 [nucl-th].

\bibitem{Leitner:2009ph}
  T.~Leitner, U.~Mosel and S.~Winkelmann,
  Phys.\ Rev.\  C {\bf 79}, 057601 (2009).

\bibitem{Anderson@NuInt09} C. E. Anderson, NuInt09 Proceedings (AIP Conf. Proc.).

\end{thebibliography}
\end{document}